\documentclass[12pt,preprint]{aastex}

\shorttitle{Circumstellar Disk}
\shortauthors{Wolf et al.}

\begin{document}


\title{The Circumstellar Disk of the Butterfly Star in Taurus}


\author{Sebastian Wolf, Deborah L. Padgett}
\affil{California Institute of Technology, 1200 E California Blvd,\\ Mail code 220-6, Pasadena, CA 91125}


\author{Karl R. Stapelfeldt}
\affil{JPL, 4800 Oak Grove Drive, Mail Stop 183-900, Pasadena, CA 91109}





\begin{abstract}
We present a model of the circumstellar environment of the so-called
``Butterfly Star'' in Taurus (IRAS~04302+2247). The appearance of this young stellar object
is dominated by a large circumstellar disk seen edge-on and the light scattering lobes above the disk.
Our model is based on multi-wavelength continuum observations:
(1) Millimeter maps, and
(2) High-resolution near-infrared obtained with {\em HST}/NICMOS.
The advantage of the combination of both observations is that they trace 
(a) different regions of the system and (b) different physical processes.
On the one hand, the millimeter-observations are sensitive to the long-wavelength radiation
being re-emitted from the dust in the central parts close to the midplane of the circumstellar disk.
Thus, the geometry and small-scale density distribution of the disk has been studied.
Furthermore, in contrast to the pure flux measurement, the resolved 1.3\,mm image allows to
discriminate between different disk models with a similar far-infrared/millimeter spectral energy distribution
and therefore to disentangle the disk geometry much more precisely.
On the other hand, the near-infrared observations trace the envelope structure and dust properties
in the envelope and the disk surface.

We find disk and envelope parameters which are comparable with those of 
the circumstellar environment of other young stellar objects.
A main result is that the dust properties must be different in the circumstellar disk
and in the envelope: While a grain size distribution with grain radii up to 100\,$\mu$m is required
to reproduce the millimeter observations of the disk, the envelope is dominated by smaller grains 
similar to those of the interstellar medium. Alternatives to this grain growth scenario
in the circumstellar disk are discussed in brief as well.
\end{abstract}




\section{Introduction}\label{intro}

The ``Butterfly Star'' in Taurus, IRAS~04302+2247, is a Class~I source located in the vicinity of Lynds~1536b
dark cloud.
Its near-infrared appearance is dominated by a totally opaque band that bisects the scattered light
nebulosity. As pointed out by Padgett et al.~1999, the apparent thickness of the extinction band decreases
by about 30\% from 1.1~$\mu$m to 2.05~$\mu$m, accounting for the reddening seen alonge its edges.
A further strong evidence on the interpretation of the dark lane as a circumstellar disk seen precisely edge-on
comes from $^{13}$CO(1-0) mapping obtained at the Owens Valley Millimeter Array (Padgett et al.~2001).
These observations indicate that the dark lane coincides with a dense rotating disk of molecular gas.

Former attempts to model the circumstellar environment of IRAS~04302+2247 have been undertaken by 
Lucas \& Roche (1997, 1998a; see also Lucas \& Roche 1998b).
The slight quadrupolar morphology of the object (therefore the alias ``Butterfly star'' - Lucas \& Roche~1997)
was assumed to be resulting from a circumstellar shell representing the late stage of contraction
of an initially spherical non-magnetic cloud combined with a cavity and an embedded opaque jet parallel to 
the axis of the angular momentum of the cloud. Based on the (later) findings of a circumstellar disk mentioned above,
we model the source as a combination of a disk plus a circumstellar envelope. The slight (mirror) symmetry,
found also in the high-resolution {\em HST}/NICMOS images of the source, is modeled under the assumption
of an optically thin outflow cone.

We base our modelling on the following observations:
(1) Resolved 1.3\,mm and 2.7\,mm maps of the disk obtained
with the {\sl Owens Valley Radio Observatory} (OVRO), and
(2) High-resolution near-infrared images obtained with {\em HST}/NICMOS at 
1.10\,$\mu$m, 1.60\,$\mu$m, 1.87\,$\mu$m, and 2.05\,$\mu$m.
While the millimeter observations are sensitive only to radiation being emitted
from dust in the dense region within the disk, the near-infrared images are dominated by
scattering of the stellar light on the dust in the circumstellar envelope and the disk
``surface''.  These observations trace different physical processes in different regions 
of the circumstellar environment, but are both strongly related to the dust properties
in the system. Thus, we are in the position not only to model both components separately
and to find the best model on the basis of common (geometrical) parameters, but
may also investigate whether the dust can be the same in the two components or not.
The latter possibility has been suggested by
investigations on the dust grain evolution in circumstellar disk where
dust grain growth alters the dust grain properties in the circumstellar disk significantly
while it is of less importance in the low-density envelope (see, e.g., Weidenschilling~1997a, 1997b).
We concentrate on modelling the circumstellar disk, and compare and verify our results with
simple models for the rather complex and highly-structured envelope.

Our millimeter and near-infrared observations and data reduction are described in brief 
in \S\ref{observations}. In \S\ref{modgen} we discuss the chosen models of both
the circumstellar disk and envelope (\S\ref{diskstruc}), define the main heating sources
of the dust (\S\ref{heat}), compile the properties of the considered dust model (\S\ref{dust}),
and give a short overview about the radiative transfer model (\S\ref{mc3d}).
Based on the millimeter maps the parameters of the circumstellar disk are derived
in \S\ref{modmm} and compared with the results of the subsequent near-infrared
light-scattering simulations in \S\ref{nircons}. Concluding remarks concerning
the interpretation of our findings are given in \S\ref{discon}.

\section{Observations}\label{observations}

\subsection{Millimeter observations}\label{ovro}

IRAS 04302+2247 was mapped in the 2.7\,mm and 1.3\,mm continua using the OVRO
Millimeter Array. The obtained maps are shown in Fig.~\ref{mmcomp} (top row).
Observations were made in four different configurations
of the six 10.4m telescopes over the period from 1998 - 2001.  
The baselines span the range 30-480 meters, and yield a beamsize of 
$1.8'' \times 1.25''$ at 2.7 mm and $0.64'' \times 0.52''$ 
at 1.3 mm using robust weighting in the AIPS IMAGR routine.  
The deconvolved source size at the two observed wavelengths is 
$2.1 \pm 0.2''$ by $1.5 \pm 0.25''$ at PA = $25\deg \pm 20\deg$ at
2.7 mm, and $2.0 \pm 0.1''$ by $0.55 \pm 0.1''$ at PA = $178\deg \pm
3\deg$ at 1.3 mm.  The measured flux densities are 22 $\pm$ 2 mJy at 
109.988 GHz and 83 $\pm$ 9 mJy at 220.399 GHz. 
The results are compiled in Tab.~\ref{mmobs}.

At the same time, we mapped IRAS~04302+2247 in $^{13}$CO 1-0 and 2-1 lines
and in the C$^{18}$O 2-1 line also at the OVRO Millimeter Array.
Strong $^{13}$CO 1-0 emission was found in a bar-like structure, 
which is centered on the position of the dust continuum source, and
is extended precisely along the position angle of the dust lane
visible in the near-infrared scattered light images 
(see \S~\ref{nicmos}, Fig.~\ref{nicmos2}) and millimeter continuum maps
(see \S~\ref{modmm}, Fig.~\ref{mmcomp}).
The gas kinematics suggest that the molecular bar is a rotating disk.
The integrated line profile is double-peaked. A velocity gradient
along the major axis of the bar with a velocity extent of 6\,km\,s$^{-1}$
has been derived. Furthermore, in the velocity channel maps
differential rotation is indicated by the shift of the $^{13}$CO emission back 
toward the central position at the velocity extrema 
(Padgett, Stapelfeldt, \& Sargent~2001). We want to remark, that
the continuum maps in combination with near-infrared images (Sect.~\ref{nicmos}) 
are the basis for the model of IRAS~04302+2247 presented in this paper, 
while the molecular line data
will be discussed in detail in a future, separate publication (Padgett, in prep.).

\begin{deluxetable}{ccccc}
\tablecaption{Millimeter observations.
  \label{mmobs}}
\tablehead{
\colhead{Wavelength}  & 
\colhead{Net Flux}    &
\colhead{Peak Flux}   &
\colhead{Beam width}  &
\colhead{Beam P.A.}   \\
\colhead{[mm]}        & 
\colhead{[mJy]}       &
\colhead{[mJy/beam]}  &
\colhead{}            &
\colhead{$[^{\rm o}]$} }
\startdata
1.36 & $83 \pm 9$ &  14.7 & $0.64'' \times 0.52''$ & -70.73\\
2.74 & $22 \pm 2$ &   8.7 & $1.8''  \times 1.25''$ & -77.81\\
\enddata
\end{deluxetable}

\subsection{Near-infrared observations}\label{nicmos}

\begin{figure}[t]
  \epsscale{1.0}
  \plotone{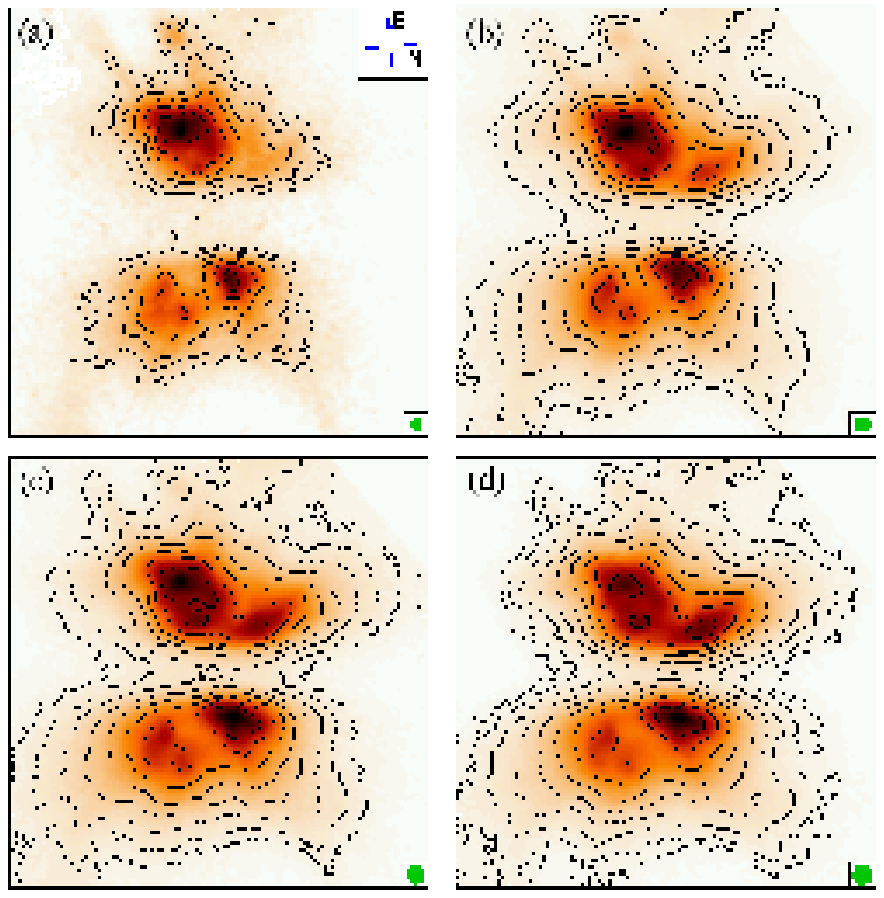}
  \caption{{\em HST}/NICMOS near-infrared scattered light images of IRAS~04302+2247:
    (a) 1.10~$\mu$m,
    (b) 1.60~$\mu$m,
    (c) 1.87~$\mu$m, and
    (d) 2.05~$\mu$m.
    Under the assumption of an object distance of 140~pc, the side length of the images is 900~AU.
    The contour lines mark steps of 0.5\,mag.
    The lowest contour lines correspond to a level of 
    4.6$\sigma$ (1.10~$\mu$m),
    9.8$\sigma$ (1.60~$\mu$m),
    8.2$\sigma$ (1.87~$\mu$m), and
    1.4$\sigma$ (2.05~$\mu$m), where $\sigma$ is the standard deviation of the background noise.
    The filled circle in the lower right edges of each image has a radius of the full width half maximum
    (FWHM) value of the point spread function (PSF): 
    0.088'' (1.10~$\mu$m), 
    0.126'' (1.60~$\mu$m),
    0.143'' (1.87~$\mu$m), and
    0.154'' (2.05~$\mu$m).  
    {\bf A preprint of this article with high-quality figures can be downloaded from: 
      http://spider.ipac.caltech.edu/staff/swolf/homepage/public/preprints/i04302.ps.gz}
  }
  \label{nicmos2}
\end{figure}

Near-infrared observations have been obtained with {\em HST}/NICMOS (see Thompson et al.~1998 for a description
of the NICMOS instrument) on August 19, 1997. NIC2 and the filters F110W, F160W,
F187W, and F205W have been used. The data reduction and flux calibration have been described in detail
by Padgett et al.~1999. The resulting scattered light images are shown in Fig.~\ref{nicmos2}.
In order not to introduce further uncertainties of the brightness distribution we did not deconvolve
the images but convolved the results of our light-scattering simulations (\S\ref{nircons}) with the
corresponding point-spread functions obtained with the program TINYTIM v.4.4 (Krist 1997).

\section{Modelling - General overview}\label{modgen}

In the following sections we give an overview about the simulation techniques and models which will be applied
in \S\ref{modmm} and \S\ref{nircons} to model the millimeter emission and near-infrared scattered light images.

\subsection{Disk and envelope structure}\label{diskstruc}

Our model of IRAS~04302+2247 consists of a circumstellar disk and an infalling envelope.
While the circumstellar disk is assumed to be responsible both for the dark lane in the optical/infrared
wavelength range and the millimeter structure of the object, an additional envelope is required to explain
the extended scattered light structure.
Although an infalling envelope, seen perpendicular to their axis of rotation symmetry, may also create a dark lane
which hides the central star at optical/infrared wavelengths (see, e.g., Lucas \& Roche~1997), 
the observed sharp transition of the dark lane towards the scattered light regions,
which could be resolved for the first time in our NICMOS/HST images, points to the presence of a disk.

For the disk we assume a density profile as described by Shakura \& Sunyaev~(1973):
\begin{equation}\label{dendisk}
\rho_{\rm disk} = \rho_0   
\left( \frac{R_{*}}{\varpi} \right)^{\alpha}
\exp \left\{ -\frac{1}{2} \left[ \frac{z}{h(\varpi)} \right]^2 \right\},
\end{equation}
where $\varpi$ is the radial distance from the star in the disk midplane, $R_{*}$ is the stellar radius,
and $h(\varpi)$ is the disk scale height:
\begin{equation}
h = h_0 \left( \frac{\varpi}{R_{*}} \right)^{\beta}.
\end{equation}
The quantities $\rho_0$ and $h_0$ will be used to scale the disk mass ($M_{\rm disk}$) 
and the scale height at a given radial distance $\varpi$ from the star, respectively.
This density profile profile has been successfully applied to model the circumstellar disks 
of HH~30~IRS (Burrows et al.~1996, Wood et al.~1998, Cotera et al.~2001) and HV~Tau~C (Stapelfeldt et al.~1998).

For the envelope we assume a density profile resulting from mass infall under consideration 
of envelope rotation (Ulrich~1976):
\begin{equation}\label{rhoenv}
\rho_{\rm env} = 
\frac{\dot{M}}{4\pi}
\left( G M_{*} r^3 \right)^{-1/2}
\left( 1 + \frac{\mu}{\mu_0}\right)^{-1/2}
\left(\frac{\mu}{\mu_0} + 2\mu_0\frac{r_{\rm c}}{r}\right)^{-1}.
\end{equation}
Here, $M_{*}$ is the mass of the central star, $\dot{M}$ is the infall rate, 
$r_{\rm c}$ is the centrifugal radius, $\mu=\cos\theta$
is the polar angle between the radial vector of the particle and the z axis, and 
$\mu_0 = f(r/r_{\rm c}, \mu)$ defines the direction of the initial streamline of the infalling material.
For $r \gg r_{\rm c}$, the density shows a steep decrease which is in agreement with 
the spectral energy distributions (SED) of
Class~I sources (see, e.g., Butner et al.~1991) and the flux distribution of Class~0 sources in
the radio wavelength range (see, e.g., Ward-Thompson et al.~1994). The density varies more slowly towards smaller
radii, achieving $\rho_{\rm env} \propto r^{-1/2}$ for $r \ll r_{\rm c}$.

\subsection{Heating sources}\label{heat}

The main heating source for the circumstellar environment is the embedded star.
Since it is completely obscured by the circumstellar disk, its effective temperature ($T_{*}$) 
and luminosity ($L_{*}$) cannot be derived directly.
The bolometric luminosity of the object ($L_{\rm bol}=0.34\,L_{\sun}$, Padgett et al.~1999) may be used
as a lower limit for the stellar luminosity. However, this is only a weak constraint since it could be derived
from the scattered light only, the amount of which is also determined by the -- rather complex -- structure
of the circumstellar envelope (see Yorke \& Bodenheimer~1999 for the ``flashlight'' effect). For a similar
reason, namely the unknown amount and optical properties of the light-scattering dust in the circumstellar envelope, 
it is too vague to derive the effective temperature of the star from the SED
of the scattered light. For these reasons we assume typical parameters of a T~Tauri star:
$R_{*} = 2\,{\rm R}_{\sun}$, $T_{*} = 4000\,{\rm K}$ (Gullbring~1998) which correspond to a luminosity of
$L_{*} = 0.92\,L_{\sun}$ under the assumption of a blackbody. As it finally turned out, 
the particular choice of a $(T_{*},L_{*})$ combination within the range of typical values
for T Tauri stars is of much less importance than 
(a) the geometry and mass of the disk and (b) the dust grain properties.

Further heating of the disk is provided by accretion, whereby we assume the source of radiation being
located in the midplane of the disk. Applying the viscous disk model by Lynden-Bell \& Pringle (1974),
the accretion luminosity $dL_{\rm acc}$ of a circular ring with the surface $dA$ and a radius $\varpi$
is given by
\begin{equation}
dL_{\rm acc} = \frac{3}{4\pi} \frac{G M_{*} \dot{M}}{\varpi^3} 
	\left[1 - \left(\frac{R_{*}}{\varpi}\right)^{1/2}\right] dA.
\end{equation}
The mass accretion rate $\dot{M}$ results from
\begin{equation}
\dot{M} = \sqrt{18\pi^3} \alpha_{\rm visc} v_{\rm c} \rho_0 \frac{h_0^3}{R_{*}},
\end{equation}
where $\alpha_{\rm visc}$ characterizes the efficiency of the mechanism of angular momentum transport
(``$\alpha$-disk'' theory, Shakura \& Sunyaev~1973) and 
\begin{equation}
v_{\rm c} = \sqrt{ \frac{G M_{*}}{R_{*}} }
\end{equation}
is the critical velocity (circular orbit).
We assume a stellar mass of $M_{*} = M_{\sun}/2$, corresponding to the other stellar parameters
given above (Gullbring et al.~1998). Furthermore, we follow Hartmann et al.~(1998), assuming an
accretion parameter $\alpha_{\rm visc}=10^{-2}$. 
In a very few cases of the most massive disks in our disk parameter space this would result in accretion
luminosities larger than observed in classical T~Tauri stars (see, e.g., Hartmann et al.~1998).
Thus we restrict the total accretion luminosity to $L_{\rm acc} \le 20\% L_{*}$ by adjusting
the parameter $\alpha_{\rm vis}$.

\subsection{Dust properties}\label{dust}

{\em Grain shape:}
We consider the dust grains to be homogeneous spheres. Although dust grains are expected to have a fractal structure,
the scattering behaviour is similar to that of spheres (see Lumme \& Rahola~1994 for porous dust particle
light scattering).
Furthermore, dust grains are expected to have a non-spherical shape 
(see, e.g., 
Elvius \& Hall~1967,
Scarrott et al.~1989,
Hajjar \& Bastien~1996,
Kastner \& Weintraub~1996,
Dollfus \& Suchail~1987,
Johnson \& Jones~1991,
Chrysostomou et al.~2000). 
However, since we do not expect the grains to be aligned on a large scale by magnetic fields and thus, 
the assumption of spherical grains is a valid approximation (Wolf et al.~2003).

{\em Grain chemistry:}
The strong absorption observed at 9.7 and 18~$\mu$m, corresponding to stretching and bending modes in silicates
on the one hand, and the strong extinction feature at 0.2175~$\mu$m which can be approximately be
reproduced by small graphite particles on the other hand (Stecher \& Donn~1965; Wickramasinghe \& Guillaume~1965),
are the basis of a widely applied grain model: It incorporates both silicate and graphite material, represented
in two separate grain populations. We use the optical data, i.e., the complex refractive index,
of ``smoothed astronomical silicate'' and graphite published by Weingartner \& Draine~2001\footnote{See also
{\tt http://www.astro.princeton.edu/$\tilde{ }$ draine}.}. Since the longest wavelength considered in our
self-consistent radiative transfer (RT) 
simulations is 2\,mm and the longest wavelength to be considered for image modelling is 2.74\,mm,
we extrapolate the refractive indexes on the basis of the astronomical silicate and graphite data published by
Draine \& Lee~(1984), ensuring a smooth transition between both data sets. 
For graphite we adopt the usual ``$\frac{1}{3}-\frac{2}{3}$'' approximation:
$Q_{\rm ext} = [Q_{\rm ext}(\epsilon_{\parallel}) + 2 Q_{\rm ext}(\epsilon_{\perp})]/3$,
where $\epsilon_{\parallel}$ and $\epsilon_{\perp}$ are the components of the graphite
dielectric tensor for the electric field parallel and perpendicular to the crystallographic $c$-axis,
respectively ($Q_{\rm ext}$ is the extinction efficiency factor). As Draine \& Malhotra (1993) have shown,
this graphite model is sufficiently accurate for extinction curve modelling.
Applying a silicate-to-graphite abundance ratio of 
$3.9\times10^{-27} {\rm cm^3} {\rm H^{-1}} : 2.3\times10^{-27} {\rm cm^3} {\rm H^{-1}}$
($R_{\rm v}$=3.1; Weingartner \& Draine~2001), we derive relative abundances of
62.5~\% astronomical silicate, 25~\% graphite ($\epsilon_{\perp}$), and 12.5~\% graphite ($\epsilon_{\parallel}$).

{\em Grain size distribution:}
We assume a grain size power-law of the form
\begin{equation} \label{gsdpl}
	n(a) \propto a^{-3.5}, \hspace*{5mm} a_{\rm min} < a < a_{\rm max},
\end{equation}
where $a$ represents the particle radius, and $a_{\rm min}$ and $a_{\rm max}$ are the minimum 
and the maximum grain radii, respectively. For $a_{\rm min}=5$\,nm and $a_{\rm max}=250$\,nm,
this grain size distribution is identical to the widely applied size distribution 
found for the interstellar medium by Mathis et al.~(1977). 
Kim et al.~(1994) showed, that the interstellar grain size distribution has to depart from a simple power law
in order to achieve a good fit to the U, B, and V extinction data where the extinction curve
changes slope and in the ultraviolet. As a solution these authors propose an additional parameter
for the size distribution to provide an exponential cutoff beyond $a_{\rm max}$.
However, in the circumstellar environment of a T~Tauri star, such as IRAS~04302+2247, the grain evolution
is expected to depend on the particular considered region:
Dust grain evolution processes, such as grain growth (see, e.g., Beckwith et al.~2000), 
are expected to occur on much shorter timescales
in the inner, dense region of the circumstellar disk than in the less dense outer regions or even the circumstellar
envelope (Weidenschilling~1997a) - thus, a radial and vertical dependence of the grain size distribution is expected. 
Furthermore, dust settling and the resulting increase of the grain--grain interaction probability  will result 
in a further vertical dependence of the grain size distribution (see, e.g., Weidenschilling~1997b). 
Beside the grain evolution, mixing processes, such as convection within the circumstellar disk have to be 
considered (see, e.g., Klahr et al.~1999).
Since the separate processes, and, even more, their mutual influence during the evolution of the circumstellar 
environment is still rather poorly understood, we assume the simple power-law given
in Eq.~\ref{gsdpl} to be valid both throughout the disk and the circumstellar envelope, 
but discuss possible corrections if necessary.
We fix the lower grain size ($a_{\rm min}=5$\,nm), but leave the upper grain size as a free parameter,
allowing us to change the dust properties, such as the opacity and scattering behaviour.

Based on the dust grain size distribution given in Eq.~\ref{gsdpl}, 
the total mass of the circumstellar disk and envelope $M$ can be estimated from
\begin{equation}
M
\propto
\eta
\rho_{\rm g}
\frac{4\pi}{3}
\left(
\frac{1-q}{4-q}
\right)
\left(
\frac{a_{\rm max}^{4-q} - a_{\rm min}^{4-q}}{a_{\rm max}^{1-q} - a_{\rm min}^{1-q}}
\right),
\end{equation}
where we assume a gas-to-dust mass ratio $\eta=100$ and a grain mass density $\rho_{\rm g}=2.5{\rm g}\,{\rm cm}^{-3}$
The quantity $q=-3.5$ is the exponent of the grain size distribution.
Since both, the gas-to-dust mass ratio and the grain mass density have no impact on the RT,
the total mass derived, e.g., for the circumstellar disk
in \S\ref{modmm} may be corrected accordingly if future observations reveal other values for these quantities.
For example, a more porous, fractal structure of the grains 
may result in a significantly smaller grain mass density $\rho_{\rm g}$ but would also have implications on
the light scattering and absorption behaviour of the grains (see, e.g., Wright~1987, Lumme \& Rahola~1994).

At this point, we would like to remind the reader, that 
only the resulting extinction and scattering behaviour of the dust grain ensemble is of importance 
for the temperature structure and reemission properties of the disk and the scattering 
in the circumstellar envelope (beside the disk geometry and density distribution 
and the heating sources). 
Since these properties of the grain ensemble are a fairly complex function of the 
grain size distribution and dust grain chemistry, each of which is based on assumptions, it is not possible
to derive a unique solution in the fitting procedure for the observed millimeter and near-infrared images.
However, restricting ourselves to a minimum set of model parameters and choosing reasonable parameters
for each of those will allow to derive at least qualitative results, while the particular quantitative
result derived from the best fit to the observed data always has to be considered in context of
the restrictions of the applied model, i.e., compact spherical homogeneous grains with a power-law
grain size distribution and the discussed silicate/graphite mixture.

The interaction between the radiation field and the dust grains is described by Mie theory.
In order to do this two problems had to be investigated first:
(1) the radiative transfer in dust grain mixtures and 
(2) the numerical description of Mie scattering in case of large size parameters.

{\em 1. Dust grain mixtures: }
For the most accurate solution of the RT problem in a dust grain mixture
one had to consider an arbitrary number of separate dust grain sizes within
a given interval $[a_{\rm min}:a_{\rm max}]$ and different chemical populations in order to derive the temperature
distribution and therefore the contribution to the (re)emitted radiation of each grain species.
Those observables, based on the RT considering each grain species
separately are close to the observables resulting from RT simulations based on weighted mean dust grain parameters
of the dust grain ensemble (Wolf~2003). 
Thus, we use weighted mean values for the efficiencies factors, cross sections,
the albedo, and scattering matrix elements. For each dust grain ensemble, $10^3$ logarithmically equidistantly
distributed grain sizes within the interval $[a_{\rm min},a_{\rm max}]$ have been taken into account for each
chemical component in the averaging process.

{\em 2. Large size parameters x: }
The size parameter $x$ is defined as the ratio of the grain radius to the wavelength of the incident 
electromagnetic radiation $\lambda$
\begin{equation}\label{grpar}
  x = \frac{2 \pi a}{\lambda}
\end{equation}
(surrounding medium: vacuum). Since previously widely applied Mie scattering codes, such as the code
by Bohren \& Huffman~(1983) fail for large size parameters ($x \gg 10^3-10^5$), we calculate the Mie scattering function
using the numerical solution for the estimation of the Mie scattering coefficients
published by Voshchinnikov~2002, which achieves accurate results both in the small as well as 
in the --~arbitrarily~-- large size parameter 
regime\footnote{{The code for calculation of the Mie scattering coefficients is available at\hfill\break}
{\tt http://spider.ipac.caltech.edu/staff/swolf/miex-web/miex.htm.}}. 
This was necessary since we consider particle sizes up to 10\,cm in the following whereby the shortest 
considered wavelength was 50\,nm, resulting in a size parameter of $x \approx 1.2 \times 10^7$.
An alternative solution would have been the combination of existing Mie scattering codes
and geometric optic solutions for the small and large size parameter range, respectively.

Although we applied the Mie scattering function for
the description of the scattering behaviour of the dust grain ensemble in the following, we also compared
the results of our simulations with those based on the much more simple approach by Heyney \& Greenstein (1941),
where the scattering parameter $g$ was derived from the scattering matrix element $F_{11}$
\begin{equation}
g = \int_0^{4\pi} \cos(\theta)\ F_{11}\frac{{\rm d}\Omega}{4\pi}
\end{equation}
and the (one-parameter) phase function can be written as
\begin{equation}
p(\cos\theta) = \frac{1-g^2}{\left[ 1 + g^2 - 2 g \cos\theta \right]^{3/2}},
\end{equation}
where $\theta$ is the scattering angle.
We found the difference between both results to be negligible compared to the influence of the other model 
parameters.

\subsection{Continuum Radiative Transfer}\label{mc3d}

For our continuum radiative transfer (RT) simulations we use the program {\em MC3D}. 
It is based on the Monte-Carlo method and solves the continuum RT problem self-consistently, i.e.,
it estimates the dust temperature distribution taking into account any heating sources.
It makes use of the temperature correction technique as described by Bjorkman \& Wood~(2001),
the absorption concept as introduced by Lucy~(1999), and the enforced scattering scheme as 
proposed by Cashwell \& Everett~(1959). The radiation field is described by Stokes parameters, i.e., 
in interaction processes of the radiation with the circumstellar matter (absorption, scattering)
its polarization state is taken into account. Multiple (anisotropic) scattering is considered.
The numerical details of {\em MC3D} have been described in previous publications 
(Wolf et al.~1999, Wolf \& Henning~2000, 
Wolf~2002\footnote{Selected parts of the {\em MC3D} documentation, results of previous applications, and 
executables of the code, prepared for particular models, are available at 
{\tt http://www.mpia-hd.mpg.de/FRINGE/SOFTWARE/mc3d/}\\ 
(current US mirror page:
{\tt http://spider.ipac.caltech.edu/staff/swolf/mc3d/}).}).
Thus, we restrict the following description on its particular adaption to the models considered 
in the subsequent sections.

In order to derive a spatially resolved dust temperature distribution, the model space has to be subdivided
into volume elements inside which a constant temperature is assumed.
Both the symmetry of the density distribution and the density gradient distribution 
have to be taken into account. We use a spherical model space, centered on the illuminating star and use a 
equidistant subdivision of the model in $\theta$-direction, while a logarithmic radial subdivision in order
to resolve the temperature gradient at the very dense inner region of the disk is applied. 
The required spatial resolution at the disk inner radius is in the order of $10^{-2}R_{*}$ 
in case of the models discussed in the subsequent chapters.

The RT is simulated at 100 wavelengths, logarithmically equidistantly distributed in the wavelength range
$[\lambda_{\rm min}, \lambda_{\rm max}] = [0.05\mu{\rm m}, 2000\mu{\rm m}]$. The complex refractive index
of the dust grain species (based on which the efficiency factors, etc., are calculated) has been derived 
by interpolation of the data sets discussed in \S\ref{dust}.

\section{Modelling: Millimeter wavelength range}\label{modmm}

\begin{figure}[t]
  \epsscale{0.6}
  \plotone{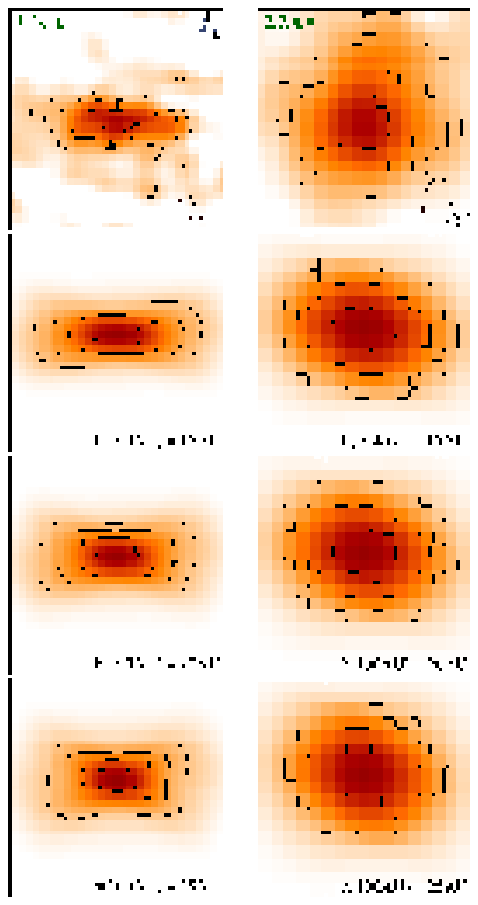}
  \caption{Comparison of the observed (upper row) and simulated (lower rows) 1.36\,mm and 2.74\,mm maps for 
    $h$(100\,AU)=15\,AU, 20\,AU, and 25\,AU. 
    The mapped region has a size of 600\,AU$\times$600\,AU.
    The ellipses in the upper images sympolize the size and orientation of the telescope beam
    (see also Tab.~\ref{mmobs} for details).
    The contour levels mark steps of 0.5\,mag, beginning at 80\% of the maximum flux.
    Further model parameters:
    $r_{\rm out}$=300\,AU,
    $\beta = 58/45$ ($\alpha=213/90 \approx 2.37$), 
    $a_{\rm max}=100\,\mu{\rm m}$,
    and
    $M_{\rm disk} = 0.07\,M_{\sun}$.}
    {\bf A preprint of this article with high-quality figures can be downloaded from: 
      http://spider.ipac.caltech.edu/staff/swolf/homepage/public/preprints/i04302.ps.gz}
  \label{mmcomp}
\end{figure}

Although the 1.36~mm and 2.74~mm images are simply structured (see Fig.~\ref{mmcomp}), 
both images provide three decisive parameters:
\begin{enumerate}
\item The net flux,
\item The peak flux, and
\item The spatial brightness distribution, whereby in case of $\lambda$=1.36~mm the spatial brightness distribution
      would even have to be described by more than one parameter.
\end{enumerate}
These parameters are found to be almost independent, i.e., within the considered parameter space,
models can be found which match one or two of these parameters but do not agree with the observations considering 
the other parameter(s). Another advantage is the fact that both the fluxes (peak, net) as well as the
spatial brightness distribution change dramatically between the $\lambda$=1.36~mm and 2.74~mm, i.e., we
have 6 (almost) independent parameters to be fitted.

Based on the model described in general in \S\ref{modgen}, we have the following adjustable model parameters:
\begin{enumerate}
\item The inner and outer disk radius ($r_{\rm in}$, $r_{\rm out}$). The inner disk radius is set to 
      $\approx 8 {\rm R}_{*}$, corresponding to an assumed dust sublimation temperature of 1600\,K, taking
      into account not only the direct stellar heating but also the reemission of the dust at the very dense,
      optically thick inner region of the disk. A first guess for the outer radius we derive from both
      the 1.36\,mm map and the near-infrared scattered light images. We assume a value of $r_{\rm out}$=300\,AU,
      but also consider models with outer radii of 200\,AU and 450\,AU.

\item The exponents $\alpha$ and $\beta$ which describe the radial density profile and disk flaring, respectively
      (Eq.~\ref{dendisk}). 
      For the flaring parameter $\beta$ we consider three different cases discussed in the literature:
      $\beta = 9/8$ (Kenyon \& Hartmann~1987),
      $\beta = 5/4$ (D'Alessio et al.~1999), and
      $\beta = 58/45$ (Chiang \& Goldreich~1997).
      We use the relation
      \begin{equation}\label{abrel}
      \alpha = 3 \left( \beta - \frac{1}{2} \right),
      \end{equation}
      which results from viscous accretion theory (Shakura \& Sunyaev~1973), to calculate a reasonable corresponding
      value of $\alpha$. These values are also in agreement with the results of former disk modelling
      efforts\footnote{The cited authors have not derived best-fit values for the parameters
	$\alpha$ and $\beta$ but have shown that the theoretically predicted values allow appropriate
	modelling of particular circumstellar disks in the T-Tauri phase.}
      (see, e.g., Burrows et al.~1996, Stapelfeldt et al.~1998, Cotera et al.~2001, Wood et al.~2002).

\item The scale height $h$ of the disk at a given radius.
      In the following we fix the scale height at a radial distance from the star of $\varpi$=100\,AU
      and consider four different cases: 
      $h$(100\,AU) = 5\,AU, 10\,AU, 20\,AU, and 40\,AU. This very wide parameter range does not only cover 
      scale heights found for other circumstellar disks but also ensures us to contain the value for IRAS~04302+2247 
      with a very high likelihood. However, a more narrow parameter range, also for the other model parameters,
      will be considered subsequently.

\item We consider disks with masses of $M_{\rm disk} = 10^{-3}, 10^{-2}$, $10^{-1}\,M_{\sun}$, and $1\,M_{\sun}$,
      corresponding to the typical masses of dust expected and found in case of T~Tauri disks 
      (Shu et al.~1987; Beckwith et al.~1990; 
      see also the publications by McCaughrean et al.~2000, Mundy et al.~2000, Natta et al.~2000,
      Wilner \& Lay~2000 and references therein).

\item For the upper radius of the grain size distribution $a_{\rm max}$ we used the following trial values:
      $a_{\rm max} = 0.25\,\mu{\rm m}$, 
      $1\,\mu{\rm m}$, 
      $10\,\mu{\rm m}$, 
      $100\,\mu{\rm m}$, 
      $1\,{\rm mm}$, 
      $10\,{\rm mm}$.
      The lowest value is justified by extinction curve modelling of the interstellar medium (Mathis et al.~1972).
      However, the slopes of the submillimeter/millimeter SEDs of circumstellar disks were found to be shallower
      than predicted from grain population of the interstellar medium (see, e.g., Beckwith et al.~1990,
      Beckwith \& Sargent~1991), indicating larger mean grain sizes: Calvet et al.~2002 suggest
      particle sizes up to $\approx$50\,cm in the presumed inner disk and $\approx$1\,cm in the outer disk
      of the 10\,Myr old protoplanetary disk of TW~Hya, and Wood et al.~2002 propose millimeter sized grains
      in the disk around the T~Tauri star HH~30. 
\end{enumerate}

The inclination of the disk was fixed at 90$^{\rm o}$ (edge-on orientation) 
since a simulation the near-infrared scattered light images
revealed an inclination of $90^{\rm o} \pm 3^{\rm o}$ (see \S\ref{nircons}).
Based on the scattered light images and the 1.3\,mm map, 
we assume a position angle of the dust lane of 175$^{\rm o}$.

For comparison with the observational results, images of the disk with the same resolution as the observed maps 
have been computed and convolved with the measured beam profile. Since the beam profiles are affected
by statistical noise, only those structures of the beam profile for which the intensity was larger 
than $3\sigma_{\rm mm}$ were taken into account, where $\sigma_{\rm mm}$ is the half width of the full maximum of 
the intensity distribution (assuming the intensity statistics of the observed map being dominated by
white background noise). For the same reason and in order to consider only the most significant features,
the net intensity measurements in case of both the observations and the simulated images
take into account values larger than 30\% of the maximum intensity.

\begin{figure}[t]
  \epsscale{1.0}
  \plotone{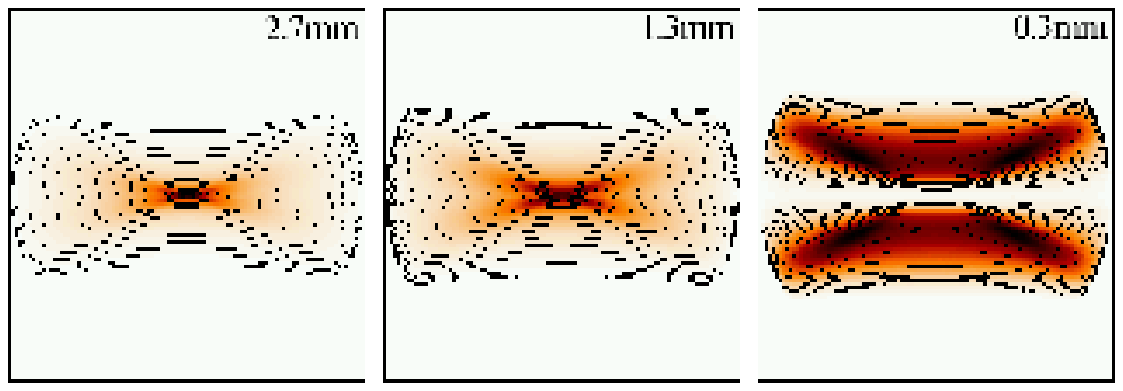}
  \caption{High-resolution intensity maps of the best fit model.
    The contour lines mark steps of 0.4 magnitudes.
    Model parameters:
    $h$(100\,AU)=15\,AU,
    $r_{\rm out}$=300\,AU,
    $\beta = 58/45$ ($\alpha=213/90 \approx 2.37$), 
    $a_{\rm max}=100\,\mu{\rm m}$,
    and
    $M_{\rm disk} = 0.07\,M_{\sun}$.}
    {\bf A preprint of this article with high-quality figures can be downloaded from: 
      http://spider.ipac.caltech.edu/staff/swolf/homepage/public/preprints/i04302.ps.gz}
  \label{hires}
\end{figure}

For grain size distributions with upper radii of $0.25\,\mu{\rm m}$, $1\,\mu{\rm m}$, and $10\,\mu{\rm m}$
the simulated net and peak fluxes are smaller than the observed fluxes.
For maximum grain sizes of $1\,{\rm mm}$ and $10\,{\rm mm}$, on the other hand, the amount of 
the measured fluxes has been achieved but no model was found for which all simulated fluxes
agree with the observed ones at the same time. One has to take into account, that the models 
for which at least separate fluxes agree with the observations are those with masses at the upper end 
of the considered mass parameter range. For those, the midplane of the disk is optically very thick 
which adds another complexity to the relation between the model parameters and the resulting fluxes,
i.e., the disk-mass to millimeter-reemission relation is far from being linear:
The higher optical depth (with increasing disk mass) results 
(a) in a less efficient heating of the disk and  
(b) in a higher absorption of the reemitted radiation (as illustrated in Fig.~\ref{hires}, the 
best-fit model discussed below is optically thick in the midplane as seen from the star
even at submillimeter/millimeter wavelengths).

\begin{figure}[t]
  \epsscale{1.0}
  \plotone{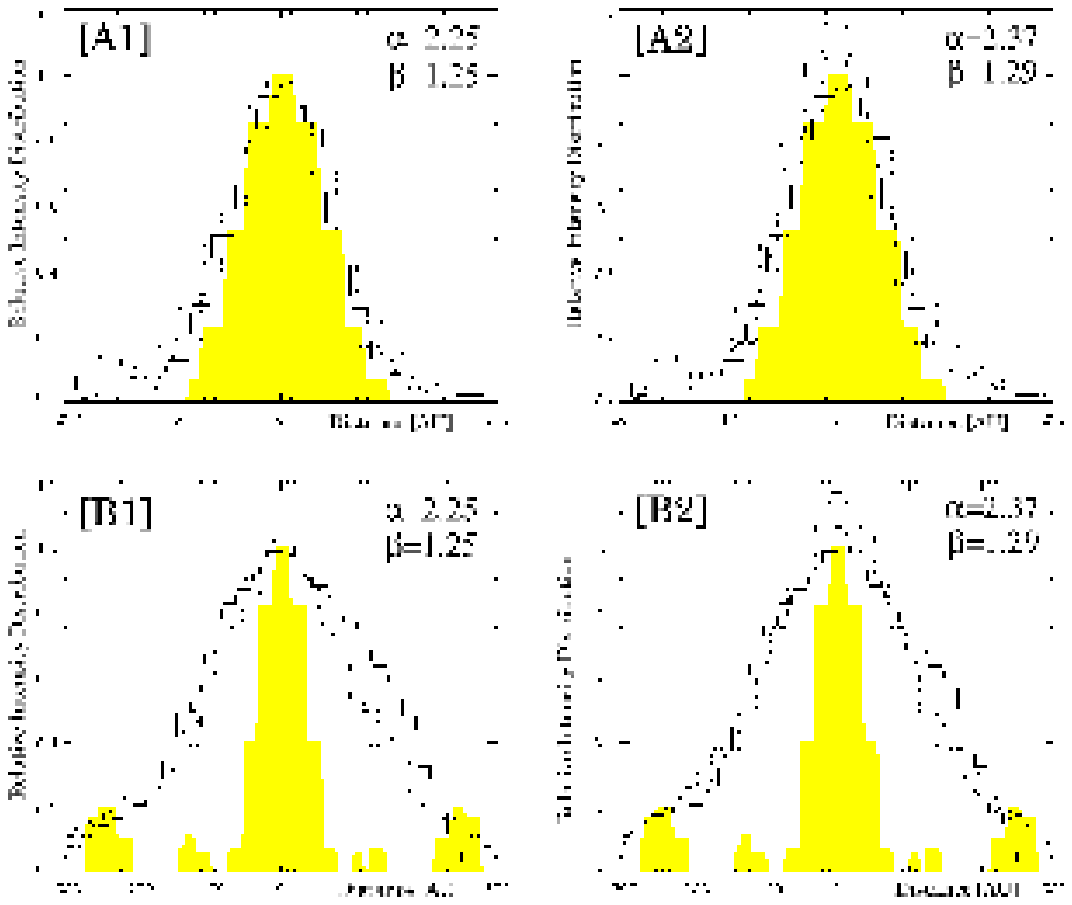}
  \caption{Relative intensity distributions centered on maximum value (1.36\,mm map).   
    {\sl [A1], [A2]:} Perpendicular to the disk midplane.
    {\sl [B1], [B2]:} Within the disk midplane.
    {\sl Solid line:} Observation, 
    {\sl Dashed line:} $h$(100\,AU)=15\,AU,
    {\sl Dotted line:} $h$(100\,AU)=20\,AU,
    {\sl Dash-dot line:} $h$(100\,AU)=25\,AU.
    Further model parameters:
    $r_{\rm out}$=300\,AU,
    $a_{\rm max}=100\,\mu{\rm m}$,
    and
    $M_{\rm disk} = 0.07\,M_{\sun}$.
    The best agreement with the observations was found for $h$(100\,AU)=15\,AU and 
    $\beta = 58/45 \approx 1.29$ ($\alpha=213/90 \approx 2.37$).
    The filled area shows the adopted beam profile along the considered direction.}
    {\bf A preprint of this article with high-quality figures can be downloaded from: 
      http://spider.ipac.caltech.edu/staff/swolf/homepage/public/preprints/i04302.ps.gz}
  \label{cuts}
\end{figure}

The only grain size distribution for which an agreement between all measured and simulated fluxes
has been found is the one with $a_{\rm max}=100\,\mu{\rm m}$. The best agreement was then found for 
$r_{\rm out}$=300\,AU,
$\beta = 5/4$ and $\beta = 58/45$, 
$h$(100\,AU)=20\,AU, and
$M_{\rm disk} = 10^{-1}\,M_{\sun}$.
In particular, the extent of the disk assuming $r_{\rm out}$=200\,AU / $r_{\rm out}$=450\,AU 
was found to be too small / too large compared with the observed image. Furthermore, all fluxes have been
found to be too small in case of $\beta = 9/8$, while their difference between models with $\beta = 5/4$ 
and $\beta = 58/45$ is negligible compared to the uncertainty of the measurements (see Fig.~\ref{cuts}
- an simultaneous increase of the $\alpha$ and $\beta$ results in a slightly more peaked intensity distribution).
Based on a more narrow parameter range around the values compiled above we found the following disk parameters:
$r_{\rm out}$=300\,AU,
$\beta = 58/45$ ($\alpha=213/90 \approx 2.37$), 
$h$(100\,AU)=15\,AU, and
$M_{\rm disk} = 0.07\,M_{\sun}$.
The optical depth in the disk midplane as seen from the star amounts to $\tau$(550\,nm)$\approx 4\times10^6$.
Thus, the disk is mainly heated by stellar radiation emitted into directions far above the midplane 
and scattered in less dense regions of the disk ($\tau \approx 1$; see Chiang \& Goldreich 1997). 

The difference of the observed and simulated peak/net fluxes
is smaller than 16\% and therefore within the error interval of the observed values.
For illustration of the results, in Fig.~\ref{mmcomp} the resulting images for three different scale heights 
are shown in comparison with the 1.36\,mm and 2.74\,mm map.
As Fig.~\ref{mmcomp} and Fig.~\ref{hires} make clear, the disk is much less resolved at 2.74\,mm
(see \S\ref{ovro} for the corresponding beam sizes).
Furthermore, Fig.~\ref{cuts} shows the intensity 
in the midplane and perpendicular to it as it has been found for $h$(100\,AU)=15\,AU, 20\,AU, and 25\,AU 
for $\beta = 5/4$ and $\beta = 58/45$, respectively.

We found further solutions for grain size distributions with maximum grain sizes larger than
investigated above (for $a_{\rm max} > 2\,$cm). However, since those models require much 
larger disk masses (which is in contradiction to T~Tauri disk models), we consider 
the described model as the most reasonable.

In Fig.~\ref{hires} 
corresponding high-resolution intensity maps at the considered wavelengths are compared with
an image at 0.3\,mm - a wavelength at which high-resolution of circumstellar disks could be obtained with
the planned Atacama Large Millimeter Array (ALMA). While the 1.3\,mm and 2.7\,mm images trace the region
close to the midplane best, high-resolution sub-millimeter maps would give even stronger constraints
on the radial disk scale height distribution than we are able to derive from the resolved,
low-resolution 1.3\,mm map.

\section{Constraints from near-infrared scattered light images}\label{nircons}

We now combine the disk and envelope by 
\begin{equation}
\rho(\vec{r}) = \left\{
	\begin{array}{r@{\quad:\quad}l}
	\rho_{\rm disk}(\vec{r}) & \rho_{\rm disk}(\vec{r}) \ge \rho_{\rm env}(\vec{r})  \\
	\rho_{\rm env}(\vec{r})  & \rho_{\rm disk}(\vec{r})  <  \rho_{\rm env}(\vec{r})  \\
	\end{array} \right. ,
\end{equation}
where $\vec{r}$ is the coordinate of any point inside the model space and 
$\rho_{\rm disk}$ and $\rho_{\rm env}$ are the density distribution of the disk (Eq.~\ref{dendisk})
and the density distribution in the envelope (Eq.~\ref{rhoenv}), respectively. 
This way, a smooth transition of the disk to the envelope density is provided without modifying the density structure
of the optically thick, millimeter-bright part of the disk\footnote{It is in principle possible to chose 
envelope parameters in Eq.~\ref{rhoenv} which result in a density in the midplane higher than that of the disk. 
However, in the considered models this is not the case.}.

Since the scattered light images reveal a highly irregularly structured envelope, it is 
the aim of the following simulations to reproduce its main parameters (but not particular details):
\begin{enumerate}
\item The wavelength-dependent width of the dust lane,
\item The relative change of the brightness distribution from 1.1~$\mu$m to 2.05~$\mu$m, and
\item The slight symmetry of the location of the brightest spots
  of the 1.60~$\mu$m to 2.05~$\mu$m images perpendicular to the midplane.
\end{enumerate}
The model is restricted to a sphere with a radius of 300\,AU around the illuminating star
in order to consider the main scattering region. We want to remark that the images in Fig.~\ref{nicmos2}
show a slight scattering also outside this range, but no further, valuable information is expected from
modelling these weak extended structures.

\begin{figure}[t]
  \epsscale{1.0}
  \plotone{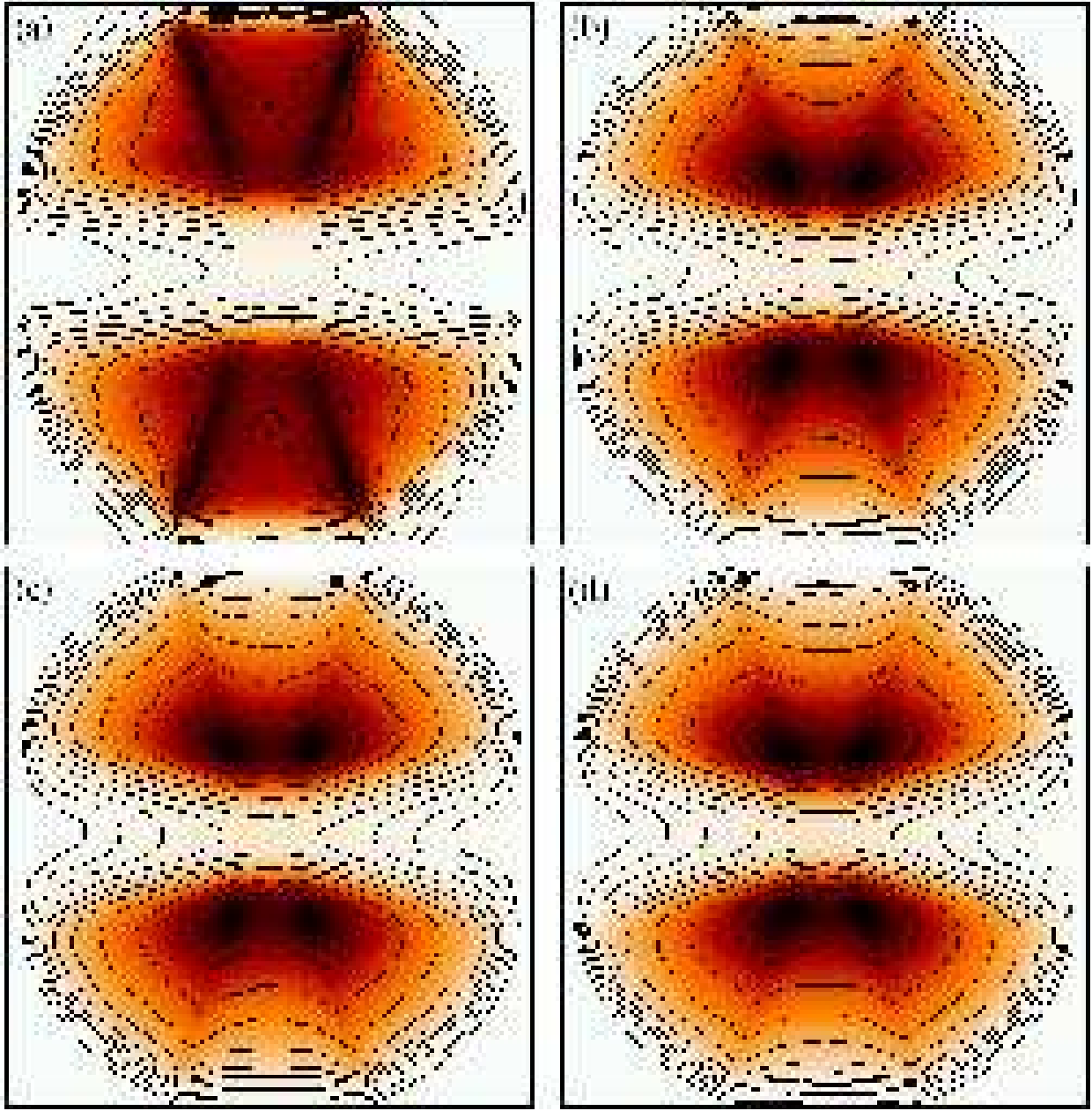}
  \caption{Near-infrared scattered light images under the assumption of small grains ($a_{\rm max}$=0.25$\mu$m)
    in the whole system (side length of the images: 600\,AU). 
    Scale height h(100\,AU)\,=\,15\,AU.
    For further disk parameters see Fig.~\ref{hires}.
    Envelope parameters: 
    $M_{\rm Disk} = 7 \times 10^{-3} M_{\sun}$,
    $\dot{M}      = 1.0 \times 10^{-7} M_{\sun}/{\rm yr}$,
    $r_{\rm c}$   = 300\,AU,
    $\mu_0        = \cos 20^{\rm o}$,
    $M_{\rm env}  = 6.1\times10^{-4}\,M_{\sun}$.
    The contour lines mark steps of 0.5\,mag.
    {\bf A preprint of this article with high-quality figures can be downloaded from: 
      http://spider.ipac.caltech.edu/staff/swolf/homepage/public/preprints/i04302.ps.gz}
  }
  \label{smallgrain1}
\end{figure}

\begin{figure}[t]
  \epsscale{1.0}
  \plotone{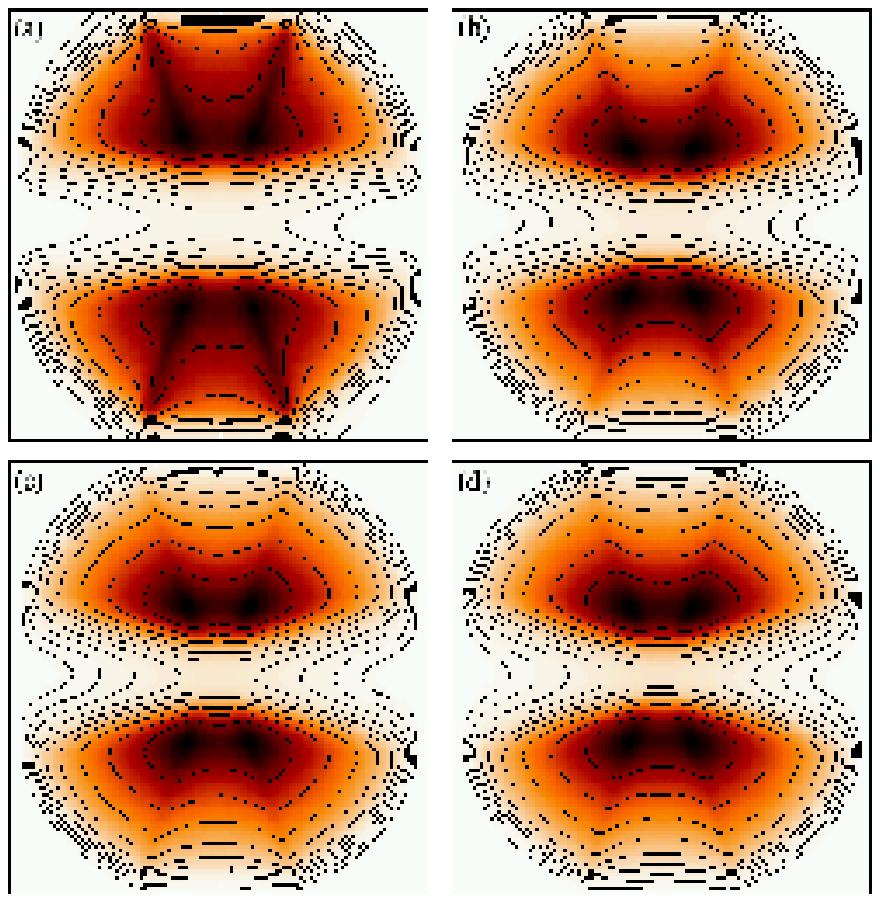}
  \caption{Near-infrared scattered light images under the assumption of small grains ($a_{\rm max}$=0.25$\mu$m)
    in the whole system (side length of the images: 600\,AU).
    Scale height h(100\,AU)\,=\,15\,AU.
    For further disk parameters see Fig.~\ref{hires}.
    Envelope parameters: 
    $M_{\rm Disk} = 7 \times 10^{-3} M_{\sun}$,
    $\dot{M}      = 0.8 \times 10^{-7} M_{\sun}/{\rm yr}$,
    $r_{\rm c}$   = 300\,AU,
    $\mu_0        = \cos 20^{\rm o}$,
    $M_{\rm env}  = 4.8\times10^{-4}\,M_{\sun}$.
    The contour lines mark steps of 0.5\,mag.
    {\bf A preprint of this article with high-quality figures can be downloaded from: 
      http://spider.ipac.caltech.edu/staff/swolf/homepage/public/preprints/i04302.ps.gz}
  }
  \label{smallgrain2}
\end{figure}

The apparent (slight) mirror symmetry of the 1.60~$\mu$m to 2.05~$\mu$m images perpendicular to the midplane
can be explained and achieved in the simulations under the assumption of optically
thin outflow cones, centered on the illuminating star. Beside geometrical arguments,
the assumption of the existence of an outflow cone is supported by a small, low-velocity
CO outflow that is associated with IRAS~04302+2247 (Bontemps et al.~1996). 
Furthermore, Gomez et al.~(1997) have found two Herbig-Haro objects several arcminutes northwest
overlying the blue-shifted lobe of the outflow.
We found the best agreement with the observations for a cone of the form
\begin{equation}
z_{\rm c}(\varpi) = \pm\left( 
z_0 + \xi \varpi^{\zeta}
\right),
\end{equation}
where $z_{\rm c}(\varpi)$ describes the boundary of the cone, 
$z_0=5$AU, $\xi=1$, and $\zeta$=1.21 ($z_{\rm c}$ and $\varpi$ are given in units of AU).
Introducing the cone one achieves 
{\sl (a)} the two intensity peaks close to the apparent upper boundary of the disk,
{\sl (b)} the increase of the intensity towards larger wavelengths, and
{\sl (c)} a ``V''-shaped increase of the intensity far above the midplane in case of 
the shorter wavelengths. While the features {\sl (a)} and {\sl (b)} are obviously ``real'',
the high intensity at shorter wavelengths high above the midplane {\sl (c)} was found only
in the upper left quadrant of the observed images. Either, the optical depth towards to
other (three) intensity peaks resulting from light scattering on the inner boundary of the outflow cone 
is higher than towards the visible one, or the only visible one is not ``real'' but
results from a local density enhancement, i.e., the ``V''-shaped intensity profile far above the midplane
is negligibly weak. The transition from the first to the second scenario can be mainly achieved 
by a decrease of the mass infall rate $\dot{M}$ and is illustrated in 
Fig.~\ref{smallgrain1} and Fig.~\ref{smallgrain2}.

\begin{figure}[t]
  \epsscale{1.0}
  \plotone{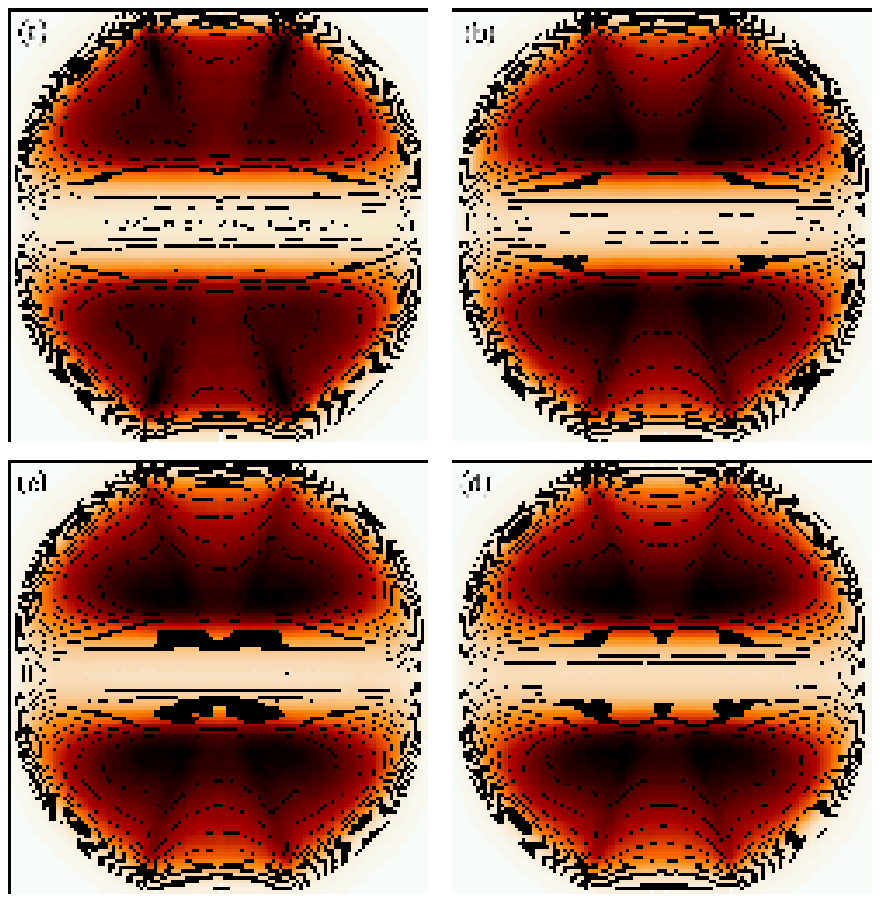}
  \caption{Near-infrared scattered light images under the assumption of large grains ($a_{\rm max}$=100$\mu$m)
    in the whole system (side length of the images: 600\,AU).
    Scale height h(100\,AU)\,=\,25\,AU.
    For further disk parameters see Fig.~\ref{hires}.
    Envelope parameters: 
    $M_{\rm Disk} = 2 \times 10^{-2} M_{\sun}$,
    $\dot{M}      = 7 \times 10^{-7} M_{\sun}/{\rm yr}$,
    $r_{\rm c}$   = 300\,AU,
    $\mu_0        = \cos 20^{\rm o}$,
    $M_{\rm env}/M_{\rm Disk}$ = 22\%.
    Since the brightness distribution is much smoother than in the observed images and
    simulations presented in Fig.~\ref{smallgrain1} and \ref{smallgrain2},
    the step width of the contour lines was chosen to be 0.25\,mag (instead of 0.5\,mag).
    {\bf A preprint of this article with high-quality figures can be downloaded from: 
      http://spider.ipac.caltech.edu/staff/swolf/homepage/public/preprints/i04302.ps.gz}
  }
  \label{large25}
\end{figure}

The width of the dust lane is mainly determined by the disk properties.
Applying the disk parameters found on the basis of the millimeter continuum measurements (\S\ref{modmm}),
the width of the dust lane is well reproduced for $\lambda=1.10\mu$m, but its decrease towards
longer wavelengths is much weaker than observed. An agreement with the observations can be achieved
under the assumption of a higher scale height $h$(100\,AU), in particular $h$(100\,AU)=25\,AU instead of
15\,AU. In this model, the mass of the disk had to be decreased by a factor of 3.5.
The resulting scattered light images are shown in Fig.~\ref{large25}.

However, this result is not satisfying since the simulated 1.36\,mm maps in Fig.~\ref{mmcomp}
clearly support a smaller scale height and a higher mass. Therefore, another parameter to be
considered is the grain size distribution in the regions being traced by the near-infrared measurements,
i.e., the envelope and the disk surface. Fig.~\ref{hires} clearly shows that the millimeter
observations trace a region which is slim compared with the width of the dust-lane in the near-infrared
wavelength range and is located below the near-infrared scattering surface of the disk.
Thus, the assumption of another dust species in the disk surface and envelope does not contradict
the 1.36\,mm/2.74\,mm model. Applying a grain size distribution as for the interstellar medium,
i.e., a minimum grain size $a_{\rm min}$=0.05\,$\mu$m and a maximum grain size $a_{\rm min}$=0.25\,$\mu$m
(Mathis et al.~1977), the near-infrared scattered light images are well reproduced under
the assumption of the geometrical disk parameters as found in \S\ref{modmm}.
If these grains are assumed in both the envelope and the whole disk, the mass of the system
amounts to $7 \times 10^{-3}\,{\rm M}_{\sun}$ which is in agreement with findings
of Moriaty-Schieven et al.~(1994) and Lucas \& Roche~(1997) who estimated a mass of the system
of $8.0 \times 10^{-3}\,{\rm M}_{\sun} \pm$ 12 per cent based on a pure envelope model without
a separate disk component (by comparison with ground-based, and therefore much less resolved
near-infrared observations), and from 800~$\mu$m photometry of the thermal dust continuum, respectively.
This mass is by a factor of 10 smaller than that resulting from our millimeter continuum model.
This huge discrepancy strongly supports the hypotheses that two distinct dust grain species are required 
to fit the millimeter and near-infrared observations simultaneously is verified.
The mass of the envelope alone amounts to about $(4.8-6.1)\times10^{-4}\,M_{\sun}$
(see Fig.\ref{smallgrain1} and \ref{smallgrain2}).
Another property of the images which can be better reproduced by the small grain size distribution 
is the overall brightness distribution:
In agreement with the observations,
the intensity is much more concentrated in the light-scattering centers on the disk surface and 
inner boundaries of the outflow cones in case of the small grain size distribution.

\begin{figure}[t]
  \epsscale{0.7}
  \plotone{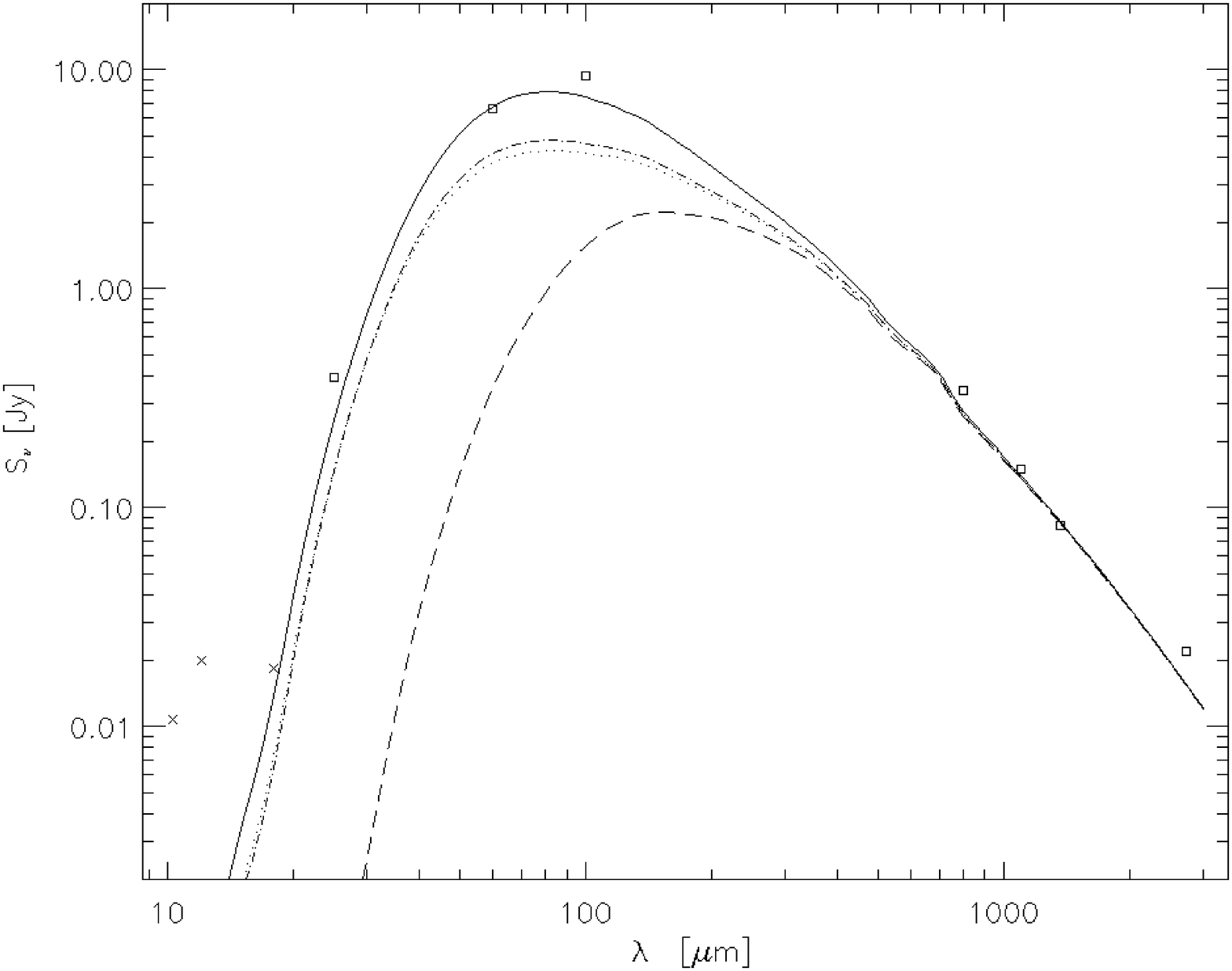}
  \caption{Dust reemission spectral energy distribution of IRAS~04302+2247.
    {\em Observations:} 
    Open squares represent observed fluxes, while crosses mark upper flux limits
    (see Tab.~\ref{obssed} for references).
    --
    {\em Simulations:}
    Dashed line:
    Pure disk re-emission (no envelope) based on the disk model found in \S~\ref{modmm}.
    Dash-dotted and Dotted line: 
    SED corresponding to the envelope(+disk) models for which near-infrared scattered light
    images are shown in Fig.~\ref{smallgrain1} and Fig.~\ref{smallgrain2}, respectively.
    Solid line:
    SED corresponding to the envelope(+disk) model presented in Fig.~\ref{smallgrain2},
    but with the envelope being extended to a maximum distance of 450~AU from the star
    above the disk midplane.
    {\bf A preprint of this article with high-quality figures can be downloaded from: 
      http://spider.ipac.caltech.edu/staff/swolf/homepage/public/preprints/i04302.ps.gz}
  }
  \label{sedsimobs}
\end{figure}

\begin{figure}[t]
  \epsscale{0.7}
  \plotone{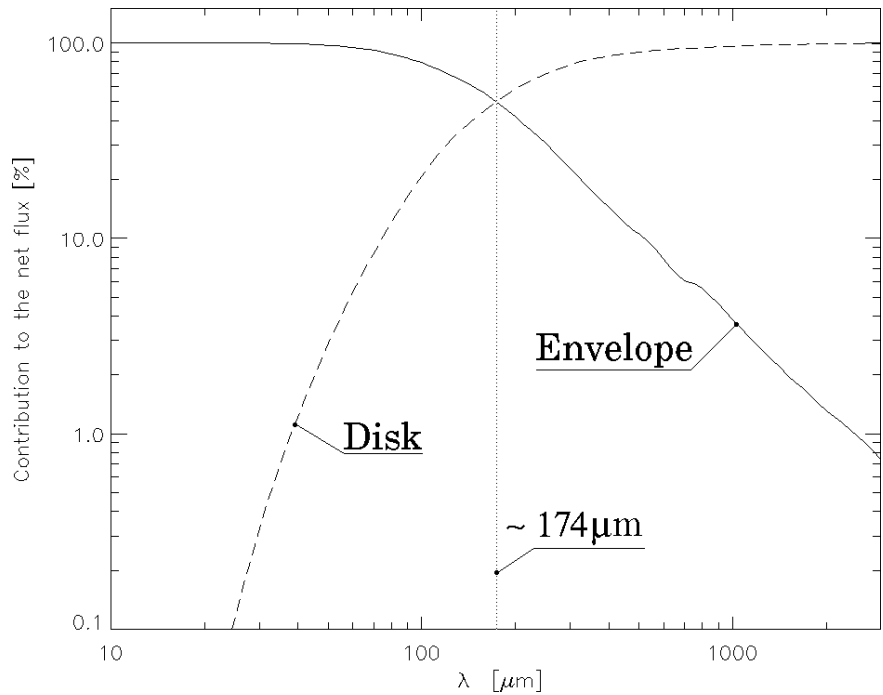}
  \caption{Separated contributions of the disk and 
    envelope to the SED in the mid-infrared to millimeter wavelength range.
    See the solid line in Fig.~\ref{sedsimobs} for the corresponding net SED.}
    {\bf A preprint of this article with high-quality figures can be downloaded from: 
      http://spider.ipac.caltech.edu/staff/swolf/homepage/public/preprints/i04302.ps.gz}
  \label{ratsed}
\end{figure}

A further, independent test of the combined envelope plus disk model can be achieved
by comparing the resulting SED with observed fluxes over a large wavelength range. 
We consider the wavelength interval $\lambda=10\,\mu$m\,-\,3\,mm where most of 
the dust reemission occurs. Scattered radiation 
in the optical to near-infrared wavelength range is not consider here, since 
the observed images show that large-scale irregular structures determine
the appearance and therefore the short-wavelength fluxes of IRAS~04302+2247.
In contrast to the scattered light images modelled above, we now also have
to consider dust outside the radius of 300~AU. According to the images 
shown in Fig.~\ref{nicmos2}, we assume a maximum extension of the envelope of 450~AU 
(radial distance to the star) above the disk. In Fig.~\ref{sedsimobs} 
the solid line as the result of our simulations shows a very good agreement 
with the observed SED. Comparing the SED of the disk alone
(dashed line in Fig.~\ref{sedsimobs}) with the reemission of the whole system
(disk+envelope), we find that the SED is dominated by the reemission
from the envelope up to a wavelength of about 174$\mu$m (see also Fig.~\ref{ratsed}).

\begin{deluxetable}{ccl}
\tablecaption{Observed Spectral Energy Distribution of IRAS~04302+2247 in the near-infrared 
  to millimeter wavelength range 
  \label{obssed}}
\tablehead{
\colhead{Wavelength}              & 
\colhead{Flux}                    \\
\colhead{[$\mu$m]}                & 
\colhead{[mJy]}                   }
\startdata
     0.90\tablenotemark{(e)}              &  $<$0.05  \\
     1.10\tablenotemark{(a)}              &     1.60  \\
     1.60\tablenotemark{(a)}              &    10.4   \\
     1.87\tablenotemark{(a)}              &    18.1   \\
     2.05\tablenotemark{(a)}              &    18.3   \\
    10.3\tablenotemark{(e)}               & $<$10.8   \\
    12\tablenotemark{(c)}                 & $<$20     \\
    17.9\tablenotemark{(e)}               & $<$18.5   \\
    25\tablenotemark{(c)}                 &   391     \\
    60\tablenotemark{(c)}                 &  6640     \\
   100\tablenotemark{(c)}                 &  9430     \\
   800 \tablenotemark{(d)}                &   342     \\
  1.1$\times10^3$\tablenotemark{(d)}      &   149     \\
  1.36$\times10^3$\tablenotemark{(b)}     &    83     \\
  2.73$\times10^3$\tablenotemark{(b)}     &    22     \\
\enddata
\tablenotetext{(a)} {This work; see \S~\ref{nicmos} for details.}
\tablenotetext{(b)} {This work; see \S~\ref{ovro}   for details.}
\tablenotetext{(c)} {IRAS flux measurements (Clark~1991).}
\tablenotetext{(d)} {Moriarty-Schieven et al.~(1994).}
\tablenotetext{(e)} {L. Hillenbrand (priv. comm.).}
\end{deluxetable}

\section{Discussion and Conclusions}\label{discon}

Based on near-infrared scattered light images and resolved millimeter maps we
have modeled the circumstellar environment of the ``Butterfly Star'' IRAS~04302+2247 in Taurus.
Our main intention was to model the properties of its large edge-on circumstellar disk for which we found
the following parameters:
\begin{itemize}
\item Outer disk radius $r_{\rm out}$=300\,AU,
\item Parameters describing the disk flaring and (relative) radial density profile:
  $\beta \approx 1.29$ and $\alpha \approx 2.37$ (whereby pre-defined $\alpha/\beta$ combinations
  have been considered), and
\item Scale height $h$(100\,AU)=15\,AU. 
\end{itemize}
The resulting mass of the disk amounts to $M_{\rm disk} = 0.07\,M_{\sun}$. 
Based on a grain size distribution $n(a) \propto a^{-3.5}$ of an interstellar-like mixture of grain
chemistries containing astronomical silicate and graphite grains we find a minimum upper grain size
of about $100\,\mu$m. While for upper grain sizes above this value up to about 2\,cm no agreement with
the millimeter observations was found, grain size distributions with maximum grain sizes above 2\,cm
do reproduce these observations as well. Nevertheless, we suggest the $100\,\mu$m solution to be
the most reasonable since larger upper grain sizes result in larger masses of the disk which is in contradiction
to theoretical models of low-mass young stellar object disks (Shu et al.~1987; Beckwith et al.~1990; 
see also the publications by McCaughrean et al.~2000, Mundy et al.~2000, Natta et al.~2000,
Wilner \& Lay~2000 and references therein).
In agreement with the results from modelling the circumstellar disks 
of HH~30~IRS (Cotera et al.~2001, Wood et al.~2002) and TW~Hya (Calvet et al.~2002) this finding
can be interpreted as grain growth in the dense regions of the disk.
However, there are two approaches which might prevent the introduction of such large grains 
in these kind of models:
(1) A flatter radial density profile, i.e., a smaller exponent $\alpha$ in Eq.~\ref{dendisk}, and 
(2) A more efficient heating 
or a higher initial temperature of the dust
in addition to heating by the embedded star. The first approach would require modifications
to already well-established circumstellar disk models, allowing the embedded star to heat
the disk midplane much more efficiently than it is possible in the $\alpha \ge 1.875$ models considered
in our investigations. We found that $\alpha$=1.2-1.5 would be required which is not met by any of these models. 
The second suggestion, however, would need a detailed re-examination of the viscous and other (also external) 
heating processes leading to a higher temperature of the dust close to the very dense regions close 
to the midplane of the disk. In agreement with the modelling efforts for, e.g., the HH~30 circumstellar disk 
mentioned above, we considered the stellar and viscous heating as the main heating sources.
But, we also found that the assumption of an initial dust temperature of about 10\,K in the dense parts
of the disk would result in the observed millimeter fluxes (and the resolved 1.3\,mm image) 
under the assumption of a smaller (e.g., the interstellar) grain size distribution as well.
Interestingly, this is the temperature regime which is assumed to be maintained in the innermost regions
of Bok globules even under the neglection of external heating sources (see, e.g., Clemens et al.~1991).

The near-infrared scattered light images in the wavelength range between 1.10\,$\mu$m and 2.05\,$\mu$m
have been used to verify the disk and dust parameters found on the basis of the millimeter observations.
It turned out that much smaller grains are required in the upper layer of the disk and
the circumstellar envelope in order to achieve an agreement between the obtained disk geometry and
the wavelength-dependent width of the dust lane between the scattering lobes. Furthermore,
the smaller grains do better reproduce the contrast between the scattering centers and the weaker structures
in the observed images.


We conclude that, based on the example of the Class~I source IRAS~04302+2247,
we found a first observational hint for different dust evolutionary scenarios in
the dense circumstellar disk and the circumstellar envelope of young stellar objects:
While dust grain growth results in $\approx 100\mu$m sized grains in the disk midplane,
the grain size in the circumstellar envelope is much more similar to interstellar medium grains.

We want to point out that further modelling of the presented observations is required
in order to verify the quantitative results:
Instead of a fixed density distribution, the vertical structure of the circumstellar disk
should be derived simultaneously with the temperature structure in an iterative procedure,
based on the condition of hydrostatic equilibrium. 
Since the accretion luminosity amounts to only 0.74\% of the stellar luminosity in
the best-fit disk model, the initial coupling between the radial and vertical density structure 
as stated by Eq.~\ref{abrel} will have to be given up
in order to find out if the theoretically predicted values for $\alpha$ and $\beta$
do really provide a unique description of the disk structure of IRAS~04302+2247.
Furthermore, the density and temperature profile at the very inner region of the massive circumstellar 
disk (within a radius of $\approx$1\,AU) has significant influence on the heating
of the outer region of the disk through dust reemission, since here the stellar radiation
-- which is the dominant energy source -- is reprocessed.
While the approximation of mean dust grain parameters describing the optical properties
of dust grain ensembles holds in the optically thick disk regime, the sublimation radii 
and dust grain temperature distribution in the inner region, 
which is directly heated by the star, strongly depend
on the particular grain size and chemical composition.
However, these questions are beyond the scope of the recent study and will be
considered in a future publication.

\acknowledgments

S.~Wolf was supported through the HST Grant GO\,9160, and through the NASA grant NAG5-11645.
We wish to thank L.A.~Hillenbrand for providing near- and mid-infrared photometric data
of IRAS~043202-2247. J.~Eisner helped to compile the flux values provided in Tab.~\ref{obssed}.

\appendix


\end{document}